\documentclass{JAC2003}

\def\ifmath#1{\relax\ifmmode #1\else $#1$\fi}%

\def\TeV{\ifmmode {\,\mathrm{ Te\kern -0.1em V}}\else
                   \textrm{Te\kern -0.1em V}\fi}%
\def\GeV{\ifmmode {\,\mathrm{ Ge\kern -0.1em V}}\else
                   \textrm{Ge\kern -0.1em V}\fi}%
\def\MeV{\ifmmode {\,\mathrm{ Me\kern -0.1em V}}\else
                   \textrm{Me\kern -0.1em V}\fi}%
\def\keV{\ifmmode {\,\mathrm{ ke\kern -0.1em V}}\else
                   \textrm{ke\kern -0.1em V}\fi}%
\def\eV{\ifmmode  {\,\mathrm{ e\kern -0.1em V}}\else
                   \textrm{e\kern -0.1em V}\fi}%

\newcommand{\ab}{\,{\rm ab}}

\newcommand{\abi}{\, \ab^{-1}}

\newcommand{\lunit}{\mathrm{cm}^{-2}\mathrm{s}^{-1}}

\newcommand{\ff}    {{\rm f}\overline{\rm f}}

\newcommand{\WW}    {\mathrm{W}^+\mathrm{W}^-}
\newcommand{\ee}    {\mathrm{e}^+\mathrm{e}^-}
\newcommand{\mumu}  {\mu^+\mu^-}

\newcommand{\bb}    {{\mathrm b\bar{\mathrm b}}}

\newcommand{\qq}    {{\rm q\bar{\rm q}}}

\newcommand{\MZ}      {m_{\mathrm{Z}}}
\newcommand{\MW}      {m_{\mathrm{W}}}
\newcommand{\MT}      {m_{\mathrm{t}}}

\newcommand {\stl}  {\sin^2 \theta_{e\!f\!f}^l}

\usepackage{graphicx}

\begin{document}

\begin{titlepage}
\vspace*{1cm}
\begin{center}

\boldmath
{\Large \bf Electroweak Gauge Theories and Alternative Theories\\
  at a Future Linear $\ee$ Collider}
\unboldmath

\vspace*{1cm}

{\sc K. M\"onig}

\vspace*{.5cm}

{\normalsize \it
DESY-Zeuthen \\
Platanenallee 6, D-15738 Zeuthen\\
email: moenig@ifh.de}
\par
\end{center}
\vskip 1cm
\begin{center}
\bf Abstract
\end{center} 
{
The measurement of Standard Model processes tests the validity of the
model at a given scale and is simultaneously sensitive to new
physics through loop effects or interference with the Standard Model
amplitudes. A variety of studies has been done to see what 
a linear collider in the energy range $\MZ < \sqrt{s} < 1 \TeV$ can offer. 
The work that has been done within the ECFA/DESY workshop on
linear colliders is reviewed, especially what was not included in
the TESLA TDR.
}
\vfill
\begin{center}
{\it Invited talk presented at the Linear Collider Workshop,
Amsterdam, April 2003}
\end{center}
\end{titlepage}
\pagestyle{plain}

\title{ELECTROWEAK GAUGE THEORIES AND ALTERNATIVE THEORIES}

\author{K. M\"onig\thanks{
The work reported in this talk was done by the members
of the ``Electroweak Gauge Theories and Alternative Theories''
working group of the Extended ECFA/DESY Study; 
B. Ananthanarayan (Bangalore),
D. Anipko (Nowosibirsk),
D. Bardin (Dubna),
I. Bozovic (VINCA Belgrade),
A. Datta (Helsinki),
A. Denner (PSI Villingen),
M. Diehl  (DESY Hamburg),
S. Dittmaier (MPI Munich),
J. Fleischer (Uni Bielefeld),
I. Ginzburg (Nowosibirsk),
J. Gluza (DESY Zeuthen),
T. Hahn (MPI Munich),
S. Heinemeyer (LMU Munich),
J. Hewett (SLAC),
W. Kilian (DESY Hamburg),
A. Leike (LMU Munich),
A. Lorca (DESY Zeuthen),
V. Makarenko (NC PHEP Minsk),
I. Marfin (NC PHEP Minsk),
M. Maul (Lund),
W. Menges (DESY Hamburg),
O. Nachtmann (Uni Heidelberg),
F. Nagel (Uni Heidelberg),
T. Ohl (Uni W\"urzburg),
P. Osland (Bergen),
A. Pankov (Gomel),
N. Paver (Trieste),
F. Piccinini (Pavia),
W. Placzek (Krakow),
P. Poulose (RWTH Aachen),
F. Richard (Orsay),
S. Riemann (DESY Zeuthen),
T. Riemann (DESY Zeuthen),
S. Rindani (Ahmedabad),
T. Rizzo (SLAC),
M. Ronan (LBL Berkeley),
M. Roth (Karlsruhe),
M. Schumacher (Uni Bonn),
J. Sekaric (DESY Zeuthen),
T. Shishkina (NC PHEP Minsk),
M. Spira (PSI Villingen),
A. Stahl (DESY Zeuthen),
D. Wackerroth (Buffalo),
A. Werthenbach (CERN),
G. Weiglein (Durham),
G. Wilson (Kansas)
},
DESY Zeuthen, Germany}

\maketitle

\begin{abstract}
The measurement of Standard Model processes tests the validity of the
model at a given scale and is simultaneously sensitive to new
physics through loop effects or interference with the Standard Model
amplitudes. A variety of studies has been done to see what 
a linear collider in the energy range $\MZ < \sqrt{s} < 1 \TeV$ can offer. 
The work that has been done within the ECFA/DESY workshop on
linear colliders is reviewed, especially what was not included in
the TESLA TDR.
\end{abstract}

\section{INTRODUCTION}
\begin{figure}[htb]
\centering
\includegraphics[width=\linewidth,bb=18 38 561 565]{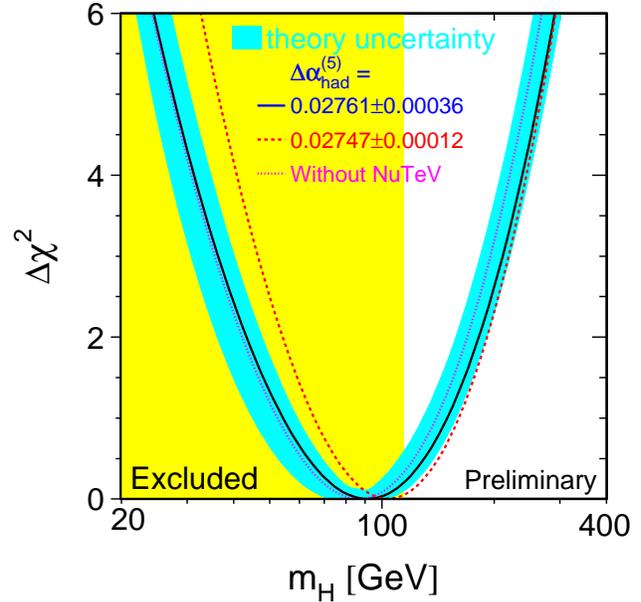}
\caption{Prediction of the Higgs mass from the electroweak precision data.}
\label{fig:blueband}
\end{figure}

\begin{figure}[htb]
\centering
\includegraphics[width=\linewidth]{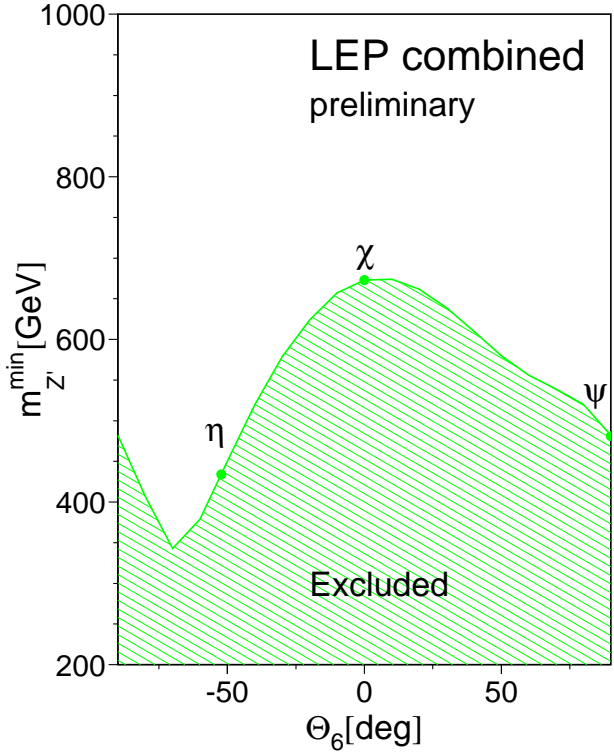}
\caption{Exclusion of Z' within E(6) models from LEP.}
\label{fig:zplep}
\end{figure}

It is a common belief that the Standard Model of electroweak interactions is
not the final theory valid up to very high scales. Nevertheless the
model is able to describe all experimental data up to now with a
typical precision around one per mille \cite{jenspdg}.
At a linear $\ee$ collider that can run at centre of mass energies, $\sqrt{s}$,
between the Z-pole and around $1 \TeV$ one expects to see finally deviations
from the Standard Model predictions.
These deviations in precision measurements occur typically for two
reasons. If the new physics occurs in loop diagrams their effect is
usually suppressed by a loop factor $\alpha / 4 \pi$ and 
very high precision is required to see it. If the new physics occurs
already on the Born level but at very high masses, the effects are
suppressed by a propagator factor 
$\frac{s}{(s-m_{NP}^2) + im_{NP}\Gamma} $ so that it is important
to work at the highest possible energies.
Both effects have already been used successfully in the past. PEP,
PETRA and TRISTAN have been able to measure the fermion couplings to
the Z although they were running at energies roughly a factor two below
the resonance pole \cite{fujii}. Ten years ago LEP could predict the
mass of the top from its loop effects \cite{leptop}, exactly 
where it was found at the TEVATRON later \cite{tevtop}. Today we are
able to limit the Higgs mass to roughly
$200\GeV$ from loop effects at LEP, SLD and the TEVATRON (figure
\ref{fig:blueband}) or to set
limits of about $500 \GeV$ on the mass of a hypothetical Z' boson 
from two fermion production
at LEP II (figure \ref{fig:zplep}) \cite{lepew}.
In the same way we expect that in ten years from now a linear
collider will estimate, depending on the physics scenario nature has
chosen, model parameters in a supersymmetric theory from
high statistics  running at the Z resonance or the mass of a 
techni-$\rho$ resonance from
W-pair production at high energies \cite{phystdr}.


There are several types of reactions to test the
Standard Model or to investigate alternative theories. With
two fermion or four fermion production on the Z-pole or close to the
W-pair production threshold one can improve on the measurements done at 
LEP and SLD by an order of magnitude.
Two fermion production at high energies is sensitive to contact
interactions in general or more specific to heavy Z'-bosons or models with
extra space dimensions.
Four or six fermion production at high energy has a large contribution
from multi gauge boson production which is sensitive to gauge boson couplings.
This is especially interesting if no elementary Higgs boson exists and the
electroweak symmetry is broken by a new strong interaction at a high scale.

In the following sections the results of the 
``Electroweak Gauge Theories and Alternative Theories'' group of the
ECFA/DESY linear collider workshop will be discussed
with particular emphasis on the progress since the TESLA TDR
\cite{phystdr} in March 2001. 

An essential ingredient for all
precision measurements are accurate Standard Model calculations which
are needed to one or two loop precision. Quite some progress has been made in
the last years and many more calculations are still under way.
This work is summarised in a special contribution to these 
proceedings \cite{stefan_lv}.

\section{THE GIGA-Z SCENARIO}

\begin{figure}[htb]
\centering
\includegraphics[width=\linewidth]{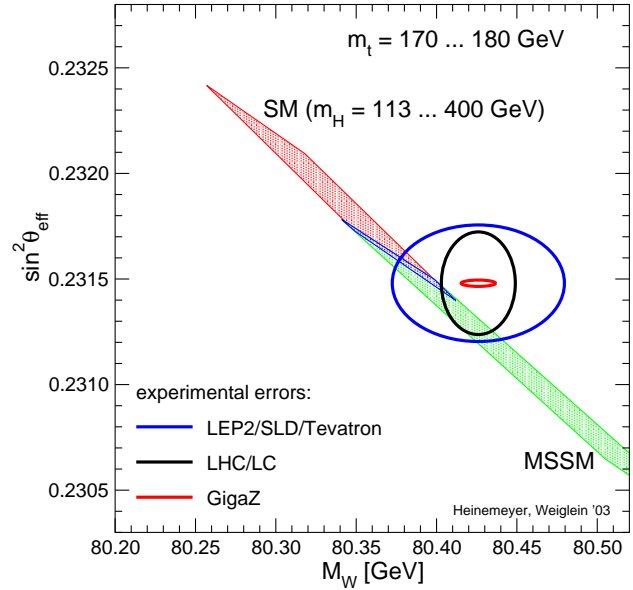}
\caption{Expected precision for $\MW$ and $\stl$ at Giga-Z compared to
  the present situation and to the LHC expectation.
  LHC/LC denotes LC high energy running only.
}
\label{fig:gigaz}
\end{figure}

The main physics goals of the Giga-Z scenario are a measurement of the
effective leptonic weak mixing angle with a precision of 
$\Delta \stl = 0.000013$ from the left-right asymmetry, which would be
an improvement of a factor 13 from LEP/SLD and a measurement of the
W-mass with an experimental accuracy of $\Delta \MW = 6 \MeV$, 
improving the present LEP/TEVATRON
result by a factor six \cite{phystdr}. While the $\stl$ measurement has no
competition at any other machine the LHC has the goal to measure the
W-mass with a precision of $15 \MeV$ \cite{lhcprec}.
The anticipated Giga-Z accuracy is shown in figure \ref{fig:gigaz} \cite{sven}
compared to the present and LHC precision and to the predictions of
the Standard Model and the MSSM.

The experimental requirements for this measurement are a luminosity of
${\cal L} \approx 5 \cdot 10^{33} \lunit$ at $\sqrt{s} \sim \MZ$ which
allows to record $10^9$ Z-decays in less than a year, electron and
positron polarisation to measure polarisation mainly from data, a beam
energy measurement of $\Delta \sqrt{s} = 1 \MeV$ relative to $\MZ$
close to the Z-peak and an extrapolation from $\MZ$ to 
$2 \MW$ with $\Delta \sqrt{s} / \sqrt{s} < 5 \cdot 10^{-5}$ 
and control of the beamstrahlung on
the few \% level. If also the Z-partial width measurements shall
be improved, an absolute measurement of the luminosity with a
precision of $\Delta {\cal L} / {\cal L} = 10^{-4}$ is needed
\cite{kmh,marc,graham}.

Excellent polarimeters are needed for relative
measurements like time dependencies or the polarisation difference between
positive and negative helicities of the beam particles. 
Detailed design studies for polarimetry, beam energy
measurement, measurement of the beamstrahlung and of the luminosity
are currently under way \cite{usbeam, lumprc}.

Significant progress was achieved on the theoretical side. The largest
parametric uncertainty for the measurement of $\stl$ is the
uncertainty in the hadronic contribution to the running of the fine
structure constant up to the Z-mass, $\alpha( \MZ^2)$. Not to be
limited too much by the knowledge of $\alpha( \MZ^2)$ the hadronic
cross section $\sigma( \ee \rightarrow \qq)$ needs to be known with 
1\% precision up to the $\Upsilon$ region \cite{fred}.
CMD II basically achieved this goal in the $\rho$ region \cite{cmd,cmd2},
however there are some discrepancies with the $\tau$ spectral
functions \cite{davier,davier2}. In the region $2\GeV < \sqrt{s} < 5 \GeV$
BES II improved the data recently from 20\% to 7\% accuracy \cite{bess}
and further
progress is possible. In addition precise results from radiative
return experiments at DA$\Phi$NE, CESR and the b-factories can be
expected in the near future.

Significant progress has been achieved also in the prediction of the
W-mass. The calculation of $\MW$ from the Fermi constant and $\MZ$ is
now complete to second order plus the $\MT$ dependent 3-loop 
corrections \cite{stefan_lv}. This results in an uncertainty in
the $\MW$ prediction of around $3-4 \MeV$.
For $\stl$ some 2-loop contributions are still missing and the
theoretical uncertainty is estimated to be $\Delta \stl = 0.00006$
much larger than the experimental goal \cite{stlerr}. 
Also some other complicated
calculations that are necessary for Giga-Z are not yet done and there
is still a long way to go.

\boldmath
\section{$\ee \rightarrow \ff$ AT HIGH ENERGY}
\unboldmath
The most general parameterisation for new physics at high scales are
contact interactions. For very large masses of the exchange particles
the propagator goes like
\[
  \left| \frac{1}{s-M^2} \right| \, \approx \, 
  \left| \frac{1}{t-M^2} \right| \, \approx \, \frac{1}{M^2}
\]
so that the new interaction can be parameterised in a contact term
$\frac{1}{\Lambda^2}$ which is equal to $\frac{g^2}{16 \pi M^2}$ in
gauge theories. 

\begin{figure}[htb]
\centering
\includegraphics[width=\linewidth,bb=13 21 472 392]{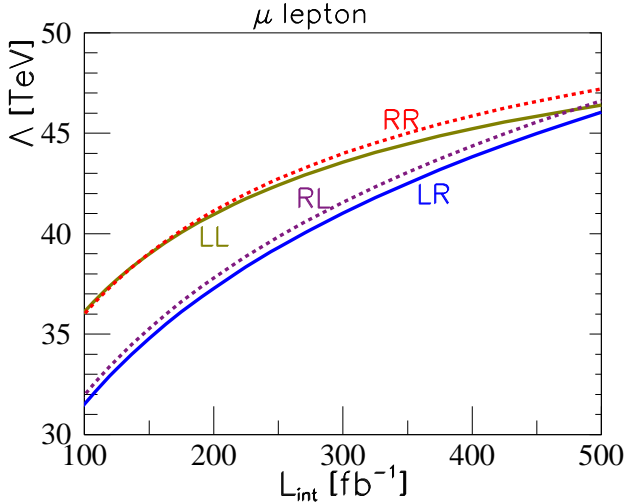}
\caption{Contact interaction reach of the linear collider
  for $\ee \rightarrow \mumu$, $\sqrt{s} = 500 \GeV$ and
  ${\cal P}_e = 0.8$ as a function of the integrated luminosity.}
\label{fig:contact}
\end{figure}

Since the
experimental sensitivity to the contact term comes mostly from the
interference with the Standard Model amplitude the helicity structure
is important. TDR studies at $\sqrt{s} = 800 \GeV$
gave typical limits around $50 \TeV$
for $\ee \rightarrow \mumu$ and $\ee \rightarrow \bb$. The LHC reaches
similar limits, however mainly for the coupling between leptons and
light quarks.
Figure \ref{fig:contact} shows the linear collider reach in $\Lambda$
for $\ee \rightarrow \mumu$, $\sqrt{s} = 500 \GeV$ and
${\cal P}_e = 0.8$ as a function of the integrated luminosity \cite{pcont}.
In a recent study the sensitivity of Bhabha and M{\/o}ller scattering to
contact interactions has been studied \cite{paver}. It was found
that the limits can be improved by typically 20\% compared to muons.
Due to the lower luminosity in ${\rm e^- e^-}$ running compared to
$\ee$ the sensitivities in Bhabha and M{\/o}ller scattering are about
the same. 

\subsection{Models with Z'}

\begin{figure}[htb]
\centering
\includegraphics[width=\linewidth]{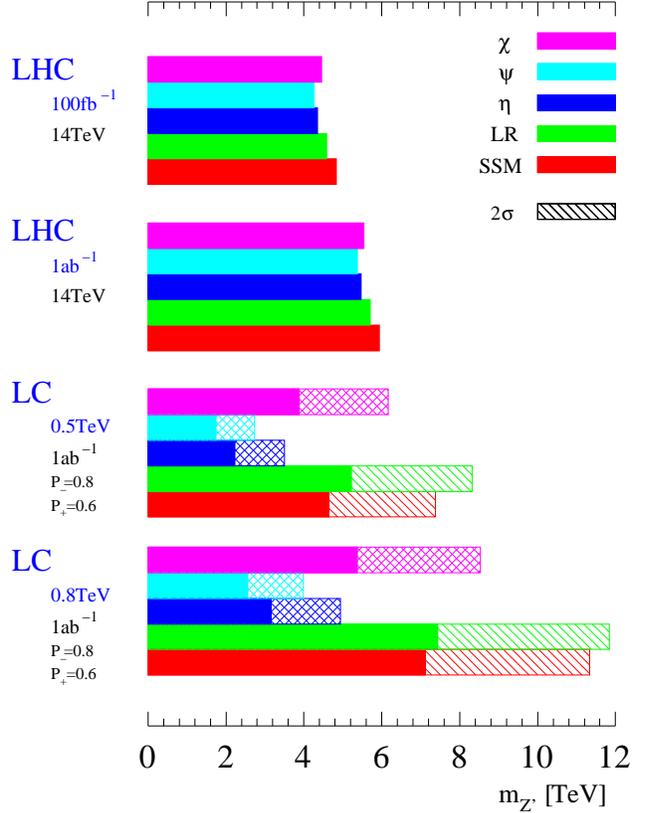}
\caption{Mass reach for a Z' in different models for LHC and LC. The
  solid bars correspond to a $5 \sigma$ discovery, while the dashed
  ones correspond to a $2 \sigma$ exclusion.}
\label{fig:zpmod}
\end{figure}

There are two approaches to study models with a Z' at a linear
collider. In a model dependent study one assumes that one knows the
model so that the Z'-mass is the only free parameter. In this case all
couplings are fixed and any
deviation of a measurement from the Standard Model value translates
directly into a value of the Z'-mass. All final states can be
used in this case. 
As for the contact terms there is a large difference between the
models since the main sensitivity comes from the interference
term. Figure \ref{fig:zpmod} compares the reachable Z' masses for
different models at the linear collider and the LHC \cite{sabine}. 
On average the reachable masses are similar for both machines and
around $4 \TeV$.

\begin{figure}[htb]
\centering
\includegraphics[width=\linewidth]{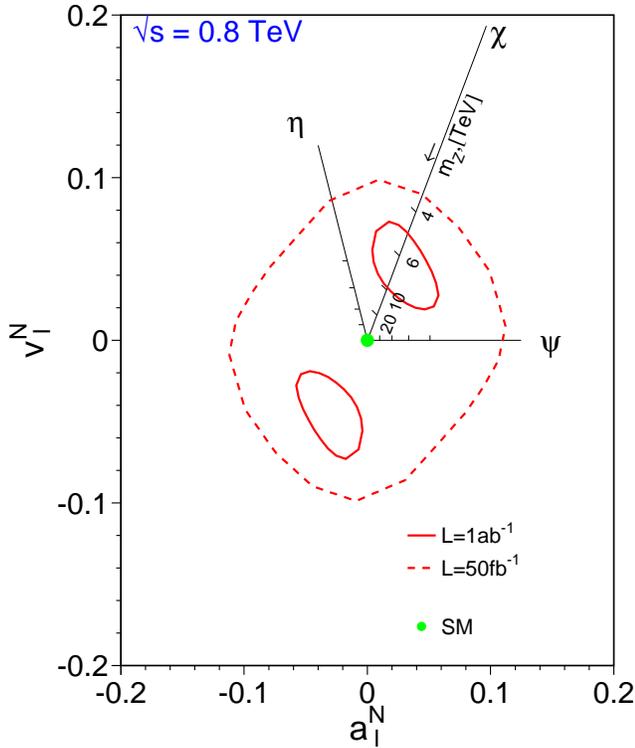}
\caption{Measurement of normalised Z' couplings at TESLA. The
  $\chi$-model with $m_{\rm Z'} = 6 \TeV$ is assumed for the central value.}
\label{fig:zpind}
\end{figure}

In a model independent approach the Z' mass and the Z' couplings are
considered simultaneously as free parameters. Any observable is given
as the product of initial state and final state couplings, so that a Z'
remains invisible in $\ee$ if it does not couple to leptons. For this
reason hadronic events can be used only when non-zero Z'-lepton
couplings are already established.
At a given centre of mass energy a linear collider is sensitive to the
normalised couplings 

\begin{eqnarray*}
  a_f^N & = & a_f' \sqrt{\frac{s}{m_{Z'}^2-s}}\\
  v_f^N & = & v_f' \sqrt{\frac{s}{m_{Z'}^2-s}}
\end{eqnarray*}
which can be measured for leptons in a model independent way using
cross sections and asymmetries. A Z' model is then defined as a line
in the $a_f^N-v_f^N$ plane where the exact position is given by the Z'
mass.
Figure \ref{fig:zpind} shows the sensitivity of the linear collider
assuming for the central value a $\chi$-model with 
$m_{\rm Z'} = 6 \TeV$, which is outside the LHC sensitivity \cite{sabine}. 
Also shown is the prediction for several E(6) models, where 
$\chi,\,\psi,\,\eta$ stands for different mixing angles between the 
gauge bosons from the ${\rm U}(1)_\chi$ and ${\rm U}(1)_\psi$ 
gauge group \cite{Cho:1998nr}.
The different models can be clearly separated with high luminosity. 

\begin{figure}[htb]
\centering
\includegraphics[width=\linewidth]{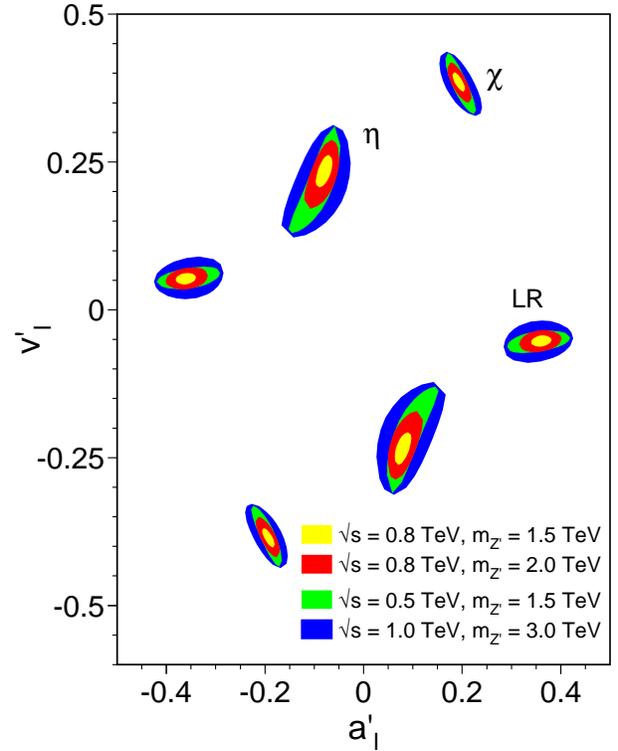}
\caption{Measurement of the Z' couplings at a linear collider for different Z'
  masses and centre of mass energies.}
\label{fig:zpcmass}
\end{figure}

In the ideal case the LHC finds a Z' and measures
its mass so the linear collider can measure the couplings. Figure
\ref{fig:zpcmass} shows the LC sensitivity in this case
for different models and
different assumptions on $\sqrt{s}$ and $m_{\rm Z'}$ \cite{phystdr}. 
In general the
couplings can be measured with a precision of a few percent.

For the left-right symmetric model an analysis on the one loop level
has been performed \cite{gluza}. In this model the quadratic top mass
dependence of $\Delta \rho$ is suppressed by a term 
$\frac{M_{W_1}^2}{M_{W_2}^2-M_{W_1}^2}$ where $W_1$ is the observed
W-boson and $M_{W_2} > 500 \GeV$. The successful prediction of $\MT$ at LEP
was therefore a pure accident and the heavy Higgs and right handed
neutrino masses need to be fine tuned to fit the LEP/SLD precision
data.

\begin{figure}[htb]
\centering
\includegraphics[width=\linewidth,bb=17 0 641 698]{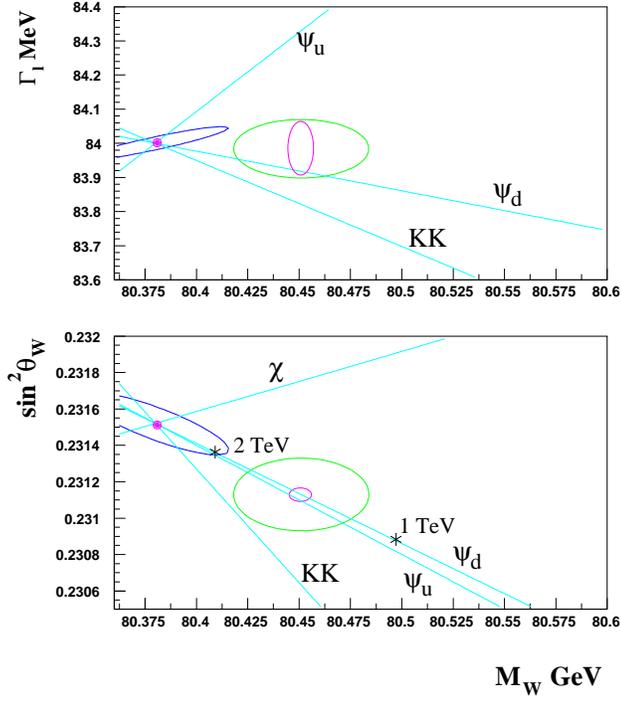}
\caption{Predictions of several models with a Z' compared to the
  present and predicted Giga-Z precision data. The ellipse around the
  crossing point indicates the uncertainty from the present day error
  on $\MT$ and $\alpha(\MZ)$.}
\label{fig:zzpmix}
\end{figure}

Another study analysed the sensitivity to Z--Z' mixing one can get
from the Z-data and the W mass \cite{richard}. As an example figure
\ref{fig:zzpmix} shows the current measurements and the Giga-Z
expectation compared to several Z' models assuming that a $115 \GeV$
Higgs has been found. It can be seen that apart from $\stl$ and $\MW$
an accurate measurement of the leptonic decay width of the Z,
$\Gamma_l$, is useful as well.

\subsection{Extra space dimensions}

The linear collider and the LHC are sensitive to the presence of 
extra space dimensions via
effects from Kaluza Klein tower graviton ($G^*$) exchange.
In the TDR it has been shown that there are visible effects from 
$\ee \rightarrow \gamma G^*$ and $\ee \rightarrow G^* \rightarrow \ff$
for an extra dimension scale of $M_H < 8 \TeV$ and $\sqrt{s} = 800
\GeV$. The reach of LHC is similar.
Recently the emphasis has been put on the question how one can
distinguish an extra dimensions signal from a Z' in case a deviation
from the Standard Model is seen. Here one can use the fact that the
Graviton is a tensor particle. 

If one defines the moments 
$\langle P_n \rangle = \int dz \frac{1}{\sigma} \frac{d \sigma}{dz} P_n(z)$, 
where
the $P_n$ are the Legendre polynomials and $z = \cos \theta$ is the
cosine of the polar angle, one can show that for vector or scalar
particle s-channel exchange  $\langle P_n \rangle=0$ for $n>2$ while for tensor
particle exchange  $\langle P_{3,4} \rangle \ne 0$ \cite{tom_mom}.
A unique identification of tensor particle exchange can be achieved up
to around $5 \TeV$ with $\sqrt{s}=800\GeV$, ${\cal L} = 1 \abi$ and
electron (positron) polarisation of 80\% (60\%). Similar results can
be obtained with specially constructed asymmetries \cite{Osland:2003fn}.

\begin{figure}[htb]
\centering
\includegraphics[width=\linewidth]{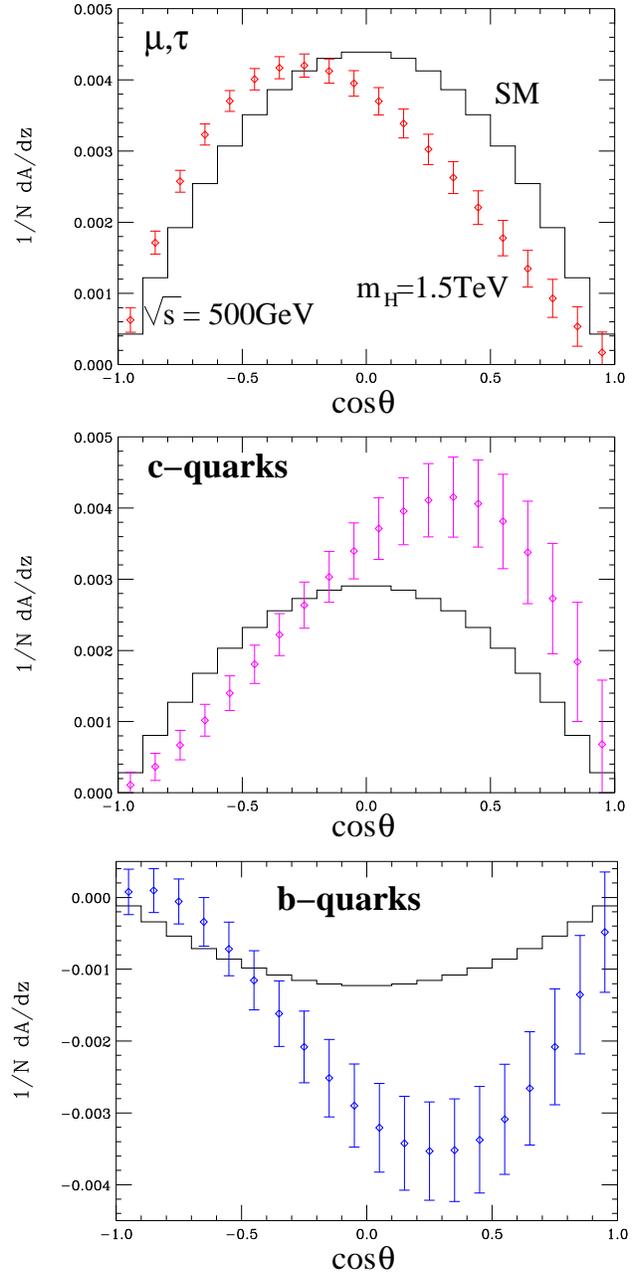}
\caption{Azimuthal asymmetry with transverse beam polarisation 
  as a function of the polar angle for
  leptons b- and c-quarks for the Standard Model and for $M_H = 1.5 \TeV$.}
\label{fig:ptasy}
\end{figure}

If transverse beam polarisation is available for both beams one can
measure the azimuthal asymmetry as a function of the polar
angle \cite{tom_pt}. 
Figure \ref{fig:ptasy} shows this asymmetry for leptons b- and
c-quarks for the Standard Model and for $M_H = 1.5 \TeV$. 
Using this asymmetry extra dimensions can be excluded up to 
$M_H< 10 (22) \TeV$ for $\sqrt{s} = 0.5 (1) \TeV$, which is the highest
reach at a next generation collider.
For vector and scalar particle exchange the azimuthal asymmetry is 
symmetric in $\cos \theta$, while it is asymmetric if also tensor 
particle exchange is
present. This allows to distinguish extra dimensions from Z' exchange
up to $M_H < 10 \sqrt{s}$.

\subsection{CP violation in $\tau$ production}
In the Standard Model the CP-violating dipole moment of the $\tau$
lepton is extremely small ($\sim 10^{-34}\,{\rm e \, cm}$). However for
example in models with Majorana neutrinos or in CP violating two Higgs
doublet models these moments can be of order $10^{-19}\,{\rm e \, cm}$.

It has been studied how well the electric and weak dipole moment can
be measured in $\tau$ pair production at TESLA using spin correlations
and polarised beams \cite{Ananthanarayan:2002fh}. 
For this analysis $\tau \rightarrow \pi \nu$ and 
$\tau \rightarrow \rho \nu$ decays have been used and CP-odd vector
correlations between the two $\tau$s have been constructed. At
$\sqrt{s}=800\GeV$ the real parts of the weak and the electromagnetic
dipole moment can be measured with a precision of 
$3-4 \cdot 10^{-19}\,{\rm e \, cm}$ which touches the
interesting region. For the imaginary parts the precision is about
three orders of magnitude worse.

\section{GAUGE BOSON PRODUCTION}

High precision measurements of gauge boson production are interesting
in several aspects. The interactions amongst gauge bosons are directly
given by the structure of the gauge group. The longitudinal gauge
bosons are connected to the mechanism of electroweak symmetry breaking
so that their interactions can teach us about this mechanism in case
no elementary Higgs boson exists.
In a strongly interacting theory the longitudinal components of 
the gauge bosons are expected to have similar interactions at very
high energies as the pions in QCD at low energy.

In a weakly interacting theory including an elementary Higgs boson 
gauge boson self-interactions should 
receive loop corrections of 
${\cal O}(\frac{g^2}{16 \pi^2}) \sim 3 \cdot 10^{-3}$. If the
experimental precision is larger than this number gauge boson
interactions should be able to test the then Standard Model in the
same way as $\stl$ and $\MW$ do it now. 

For the TDR a detailed study including full detector simulation has
been done for $\ee \rightarrow \WW$ \cite{menges}. It has been found
than the C, P conserving couplings can be measured with a precision of 
$3-15 \cdot 10^{-4}$ at $\sqrt{s} = 500 \GeV$ and around a factor two
better at $\sqrt{s} = 800 \GeV$. This is much better than the
expected effects from radiative corrections so that W-pair production
will become a new precision observable. In a strongly interacting
scenario this precision translates into a new physics scale of $\Lambda > 10
\TeV$ which is also significantly above the $\Lambda \le 3 \TeV$ limit 
from unitarity. The C or P violating couplings can be measured roughly
one order of magnitude worse than the C, P conserving ones.

Recently a study using optimal observables has been done
\cite{diehl}. This work is on parton level only up to now, but it has
shown that the imaginary parts of the couplings can be measured
simultaneously with the real parts with about the same precision and
without degrading the precision of the real parts. Only one
combination of couplings $({\rm Im}(g^R_1+ \kappa_R))$ cannot be
measured with longitudinal beam polarisation. 
If transverse beam polarisation is available this coupling can be measured.
In this case, however, the precision of the other coupling is degraded by 
roughly a factor of two \cite{diehl2}.

\begin{figure}[htb]
\centering
\includegraphics[width=\linewidth]{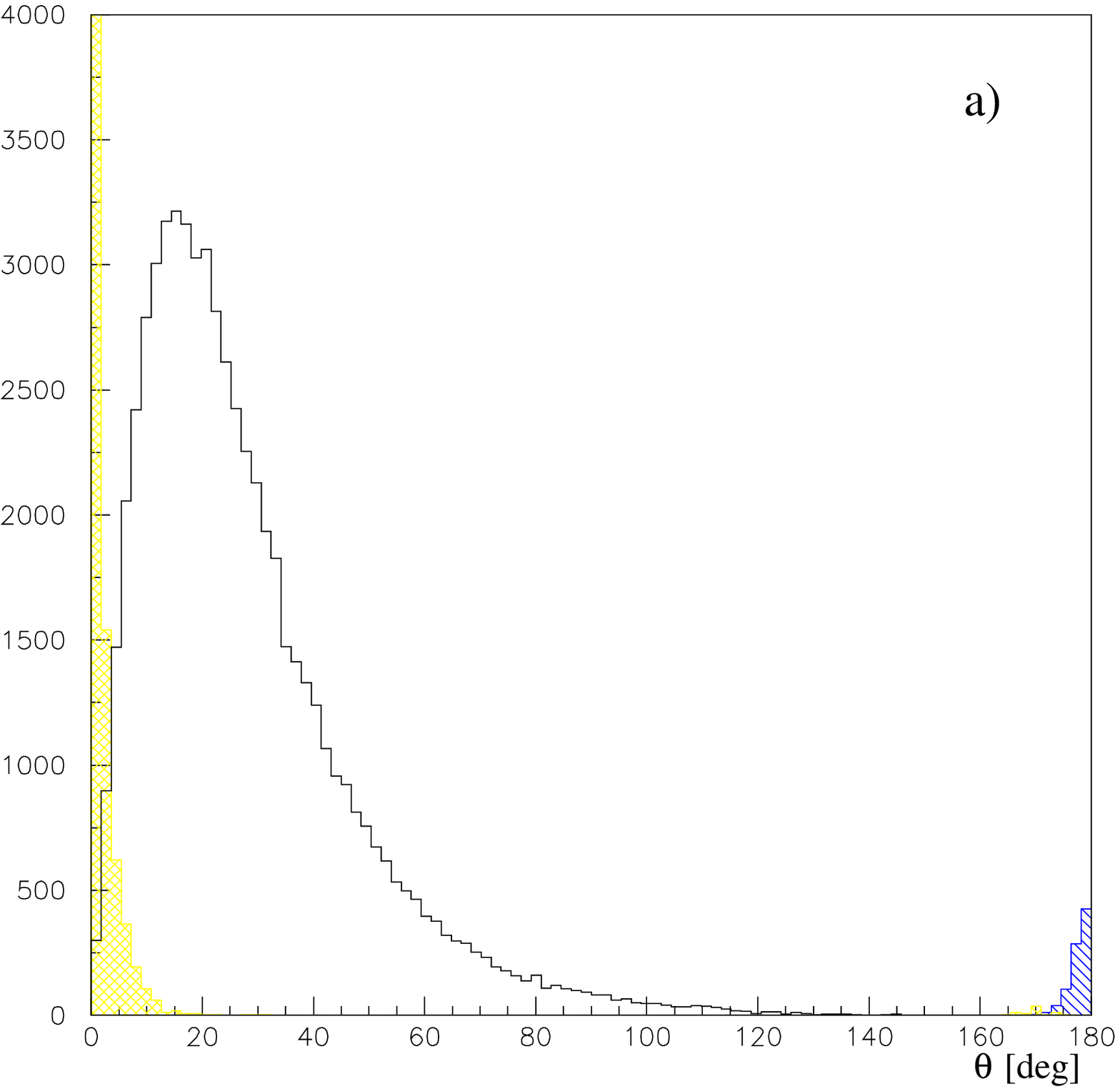}
\includegraphics[width=\linewidth]{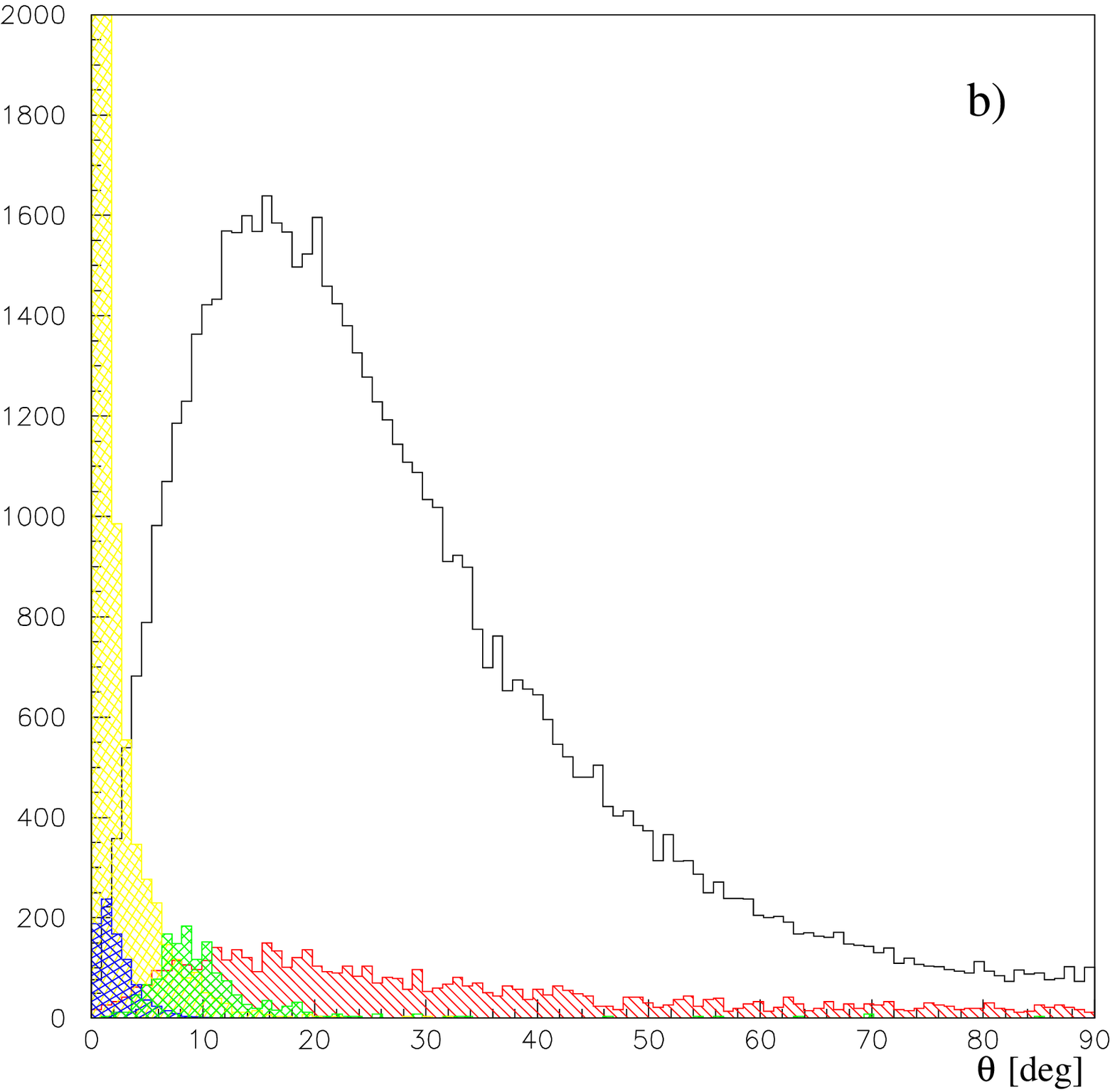}
\caption{Signal and background for $e \gamma \rightarrow W \nu$ in the
  real $e \gamma$ mode (a) and from the parasitic $\gamma \gamma$ running (b).
  The white area represents the signal. In a) the hatched contribution
  on the left is from $\gamma \gamma$ induced processes and the one on
  the right from $e \gamma \rightarrow eZ$. The additional cross-hatched 
  (green) contribution in b) is from 
  $\gamma \gamma \rightarrow q \bar{q}$ and the singly hatched (red) from 
  $\gamma \gamma \rightarrow WW$.
}
\label{fig:egwn}
\end{figure}

Also the measurement of the triple gauge couplings at a photon collider in 
$\gamma \gamma \rightarrow WW$ and $e \gamma \rightarrow W \nu$ has
been studied, using hadronic W decays only. 
The study of $e \gamma \rightarrow W \nu$ is reasonably
complete \cite{jadranka}, while in $\gamma \gamma \rightarrow WW$ the
azimuthal decay angle, $\phi$, which is sensitive to the interference
of the different helicity amplitudes is still missing \cite{ivanka}.
Both reactions can be selected cleanly with an overall efficiency
around 80\%. Figure \ref{fig:egwn} shows the polar angle distribution
for $e \gamma \rightarrow W \nu$ and the background after cuts on the
visible energy and the invariant mass. In the real $e \gamma$ mode,
where only one beam is converted, only some background in the extreme
forward and backward regions is left from $e \gamma \rightarrow Ze$
and from $\gamma \gamma$ induced hadron production which can easily
be rejected by an angular cut. In the parasitic mode, where the 
$e \gamma$ luminosity during $\gamma \gamma$ running is used, some
additional background from $\gamma \gamma \rightarrow WW$ where one W
decays leptonically is left.

\begin{figure}[htb]
\centering
\includegraphics[width=\linewidth,bb=33 0 491 468]{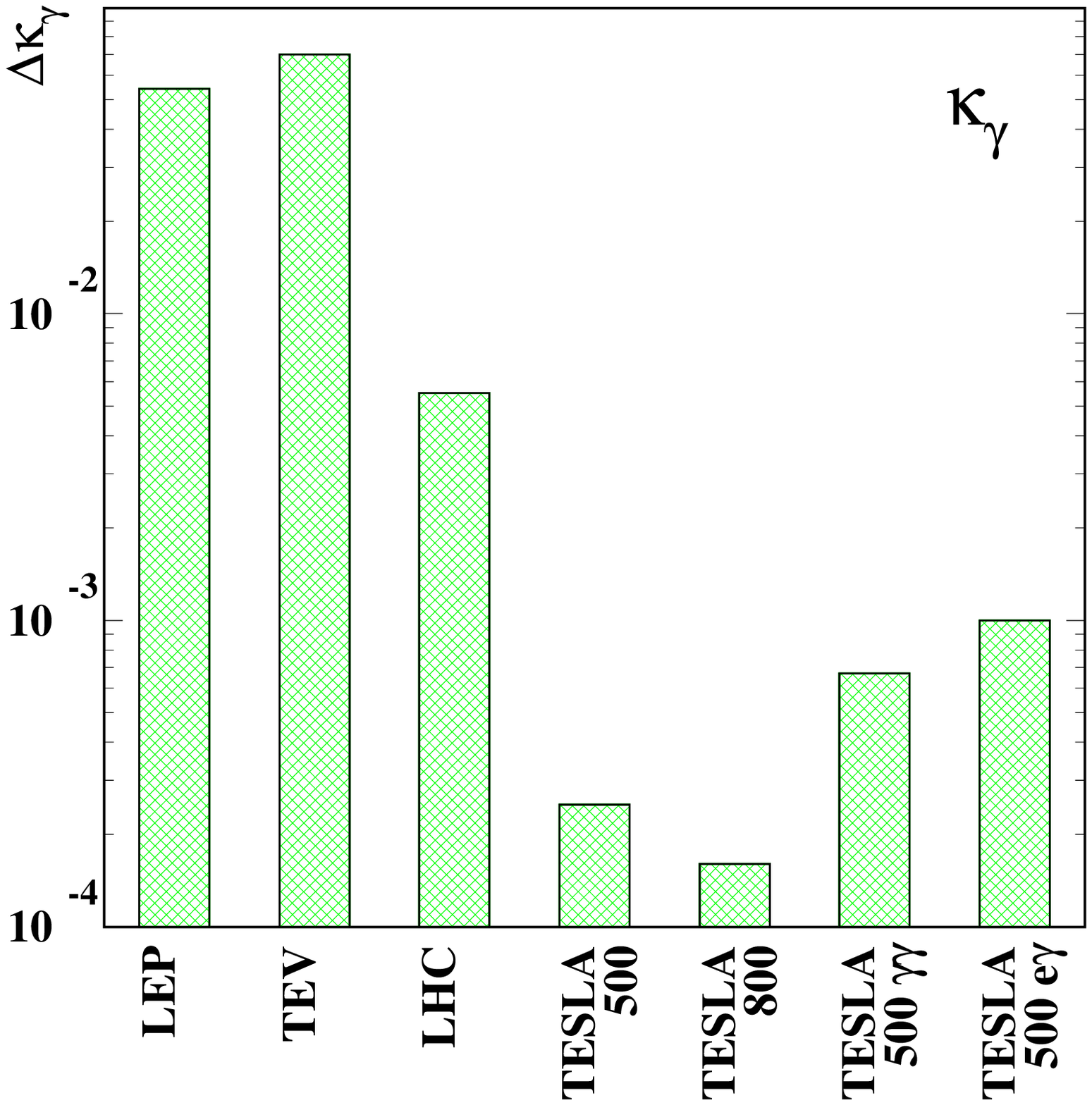}
\includegraphics[width=\linewidth,bb=33 0 491 468]{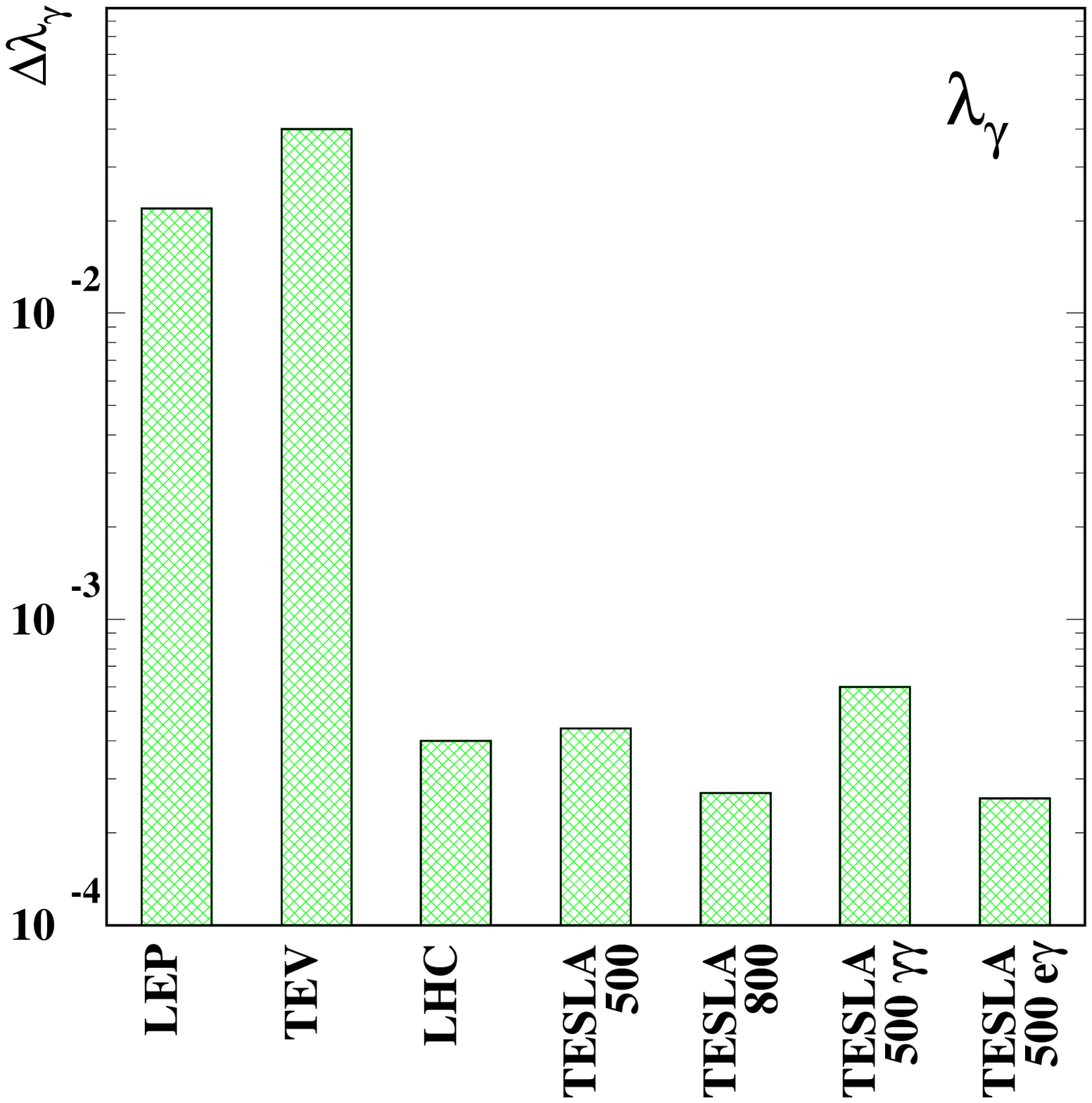}
\caption{The expected precision for $\kappa_\gamma$ and
  $\lambda_\gamma$ at different machines.}
\label{fig:tgc}
\end{figure}

The cross sections in these two channels are much larger than in
$\ee$. However there are no large gauge cancellations so that the
final precision is comparable in all cases.
Figure \ref{fig:tgc} compares the expected precision for
$\kappa_\gamma$ and $\lambda_\gamma$ at the different machines. For
$\gamma \gamma$ and $e \gamma$ a 0.1\% error on the luminosity is assumed.
It should be noted that $\kappa_\gamma$ is very sensitive to the
luminosity error and to uncertainties in the polarisation while
$\lambda_\gamma$ is basically insensitive to these effects. For 
$e \gamma$ the improvement using the $\phi$ angle in the fit is a
factor seven for $\lambda_\gamma$, a similar factor can be expected for
$\gamma \gamma$ as well. In summary $\kappa_\gamma$ will be measured
significantly worse in $e \gamma$ and $\gamma \gamma$ than in $\ee$,
however still good enough for cross checks in case deviations from the
Standard Model are found. For $\lambda_\gamma$ the photon collider
could give the best result.

In an alternative study the leptonic W decays in $e \gamma \rightarrow W \nu$
have been considered \cite{anipko}. In these events only a single lepton is
seen in the detector. The couplings have been measured from the cross section
in an optimised phase space region where background and the variable photon
energy has been taken into account. Assuming no error on the normalisation,
the error in $\kappa_\gamma$ is similar for the two analyses taking the
lower leptonic branching ratio of the W into account. For $\lambda_\gamma$
the error in the leptonic analysis is significantly larger because of the
missing information due to the second missing neutrino.

\begin{figure}[htb]
\centering
\includegraphics[width=\linewidth]{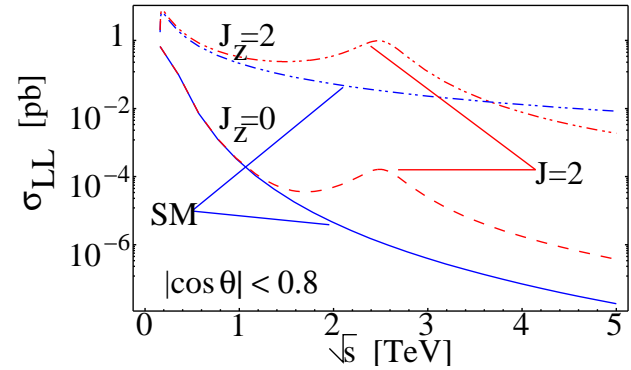}
\caption{Cross section for longitudinal W-pair production in $\gamma
  \gamma$ scattering for the Standard Model and in presence of a
  tensor resonance with $2.5 \TeV$ mass. $J_z$ denotes the spin of
  the $\gamma \gamma$ system.}
\label{fig:ggre}
\end{figure}

It is known since long that $\ee \rightarrow \WW$ is sensitive to
technicolour vector resonances in the same way as the $\rho$ is seen
in $\ee \rightarrow \pi^+ \pi^-$ \cite{tim}. 
It has been shown now, that 
$\gamma \gamma \rightarrow \WW$ is sensitive to rescattering effects
from a scalar or a tensor resonance \cite{poulose}. Figure
\ref{fig:ggre} compares the cross section for longitudinal gauge boson
production in the central region for the Standard Model and for a
tensor resonance with a mass of $2.5 \TeV$. An experimental study, if
these effects are measurable at TESLA, is planned. 

These studies underline
the importance to measure the gauge couplings in several different
channels. 
For example a vector resonance would result in anomalous gauges
couplings in $\ee$ while in $\gamma  \gamma$ and $e \gamma$ one might
still measure the Standard Model values.

\begin{figure}[htb]
\centering
\includegraphics[width=\linewidth]{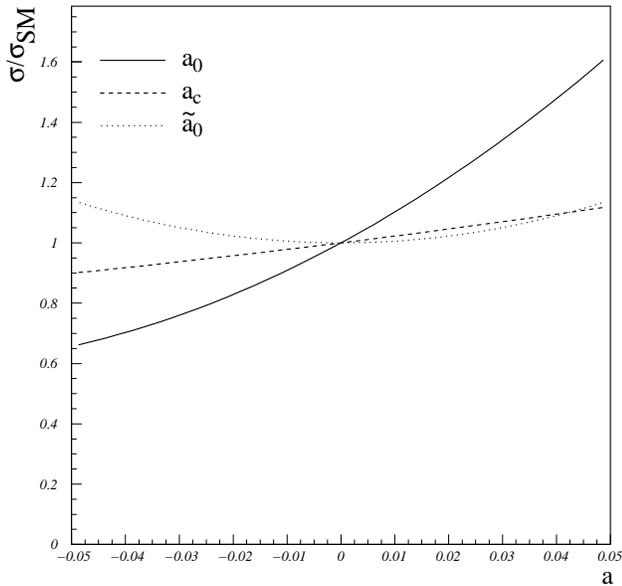}
\caption{Dependence of $\sigma(\gamma \gamma \rightarrow \WW)$ for
  $\sqrt{s} = 1 \TeV$ on the $\gamma \gamma \WW$ couplings. The exact
  definition of the couplings can be found in \cite{stefan,wwgg}.}
\label{fig:wwgg}
\end{figure}

The reaction $\gamma \gamma \rightarrow \WW$ is also the ideal place
to test for anomalous $\gamma \gamma \WW$ quartic couplings. 
These couplings have been first studied in $\ee \rightarrow \WW \gamma$
and limits of the coupling parameters of ${\cal O}(1)$ at $\sqrt{s}=500\GeV$
have been found \cite{anja,stefan}.
The cross section
dependence of $\gamma \gamma \rightarrow \WW$ on these couplings has
been studied and limits on these couplings have been derived \cite{wwgg}.
Figure \ref{fig:wwgg} shows the cross section dependence on these
couplings for $\sqrt{s} = 1 \TeV$. Without systematic uncertainties
limits between $10^{-4}$ and $10^{-2}$ can be achieved. This is about
three orders of magnitude better than the $\ee$ result.

\section{CONCLUSIONS}
It has been shown that electroweak precision tests contribute
significantly to the physics of a linear collider.
Precision measurements on the Z pole can test model parameters inside or
beyond the Standard Model.
Two-fermion production at high energy tests a wide class of models
like those containing additional Z' bosons or extra space dimensions. 
The limits are often comparable or better than those at the LHC. 
W-pair production provides new precision observables on the same level 
as $\stl$ or $\MW$. If no light Higgs exists, gauge boson production 
offers a window to strong electroweak symmetry breaking.

In summary it is the combination of the direct studies of the probable 
extensions of the
Standard Model, like Higgs and SUSY, with the potential of the precision tests 
that makes the Linear Collider a unique tool to understand the physics of 
electroweak symmetry breaking.

\end{document}